\documentclass{INTERSPEECH2023}

\usepackage{url,hyperref, amsmath,booktabs, multirow, graphicx, xcolor}
\usepackage{epstopdf}

\interspeechcameraready

\title{The DISPLACE Challenge 2023 - DIarization of SPeaker and LAnguage in Conversational Environments}

\name{Shikha Baghel${}^1$, Shreyas Ramoji$^1$, Sidharth$^1$, Ranjana H$^1$,  Prachi Singh$^1$, Somil Jain$^2$, Pratik Roy Chowdhuri$^2$,  Kaustubh Kulkarni$^1$, Swapnil Padhi$^1$, Deepu Vijayasenan$^2$, Sriram Ganapathy$^1$
 \thanks{This work was funded by the project National Language Translation Mission (NLTM): BHASHINI, the Ministry of Electronics and Information Technology (MeitY), Government of India SP/MITO-22-001 grant.}
\address{${}^1$LEAP Lab, Department of Electrical Engineering, Indian Institute of Science, Bengaluru, India
  ${}^2$Department of Electronics and Communication, National Institute of Technology Karnataka,\\ Surathkal, India}
\email{shikhabaghel@iisc.ac.in}}

\begin{document}
\maketitle
\begin{abstract}
In multilingual societies, social conversations often involve code-mixed speech. The current speech technology may not be well equipped to extract information from  multi-lingual multi-speaker conversations. The DISPLACE challenge entails a first-of-kind task to benchmark speaker and language diarization on the same data, as the data contains multi-speaker conversations in multilingual code-mixed speech. The challenge attempts to highlight outstanding issues in speaker diarization (SD) in multilingual settings with  code-mixing. Further, language
diarization (LD) in multi-speaker settings also introduces new challenges, where the system has to disambiguate speaker switches with code switches. For this challenge, a natural multilingual, multi-speaker conversational dataset is distributed for development and evaluation purposes. The systems are evaluated on single-channel far-field recordings. We also release a baseline system and report the highlights of the  system submissions.
\end{abstract}

\noindent\textbf{Index Terms}: Speaker diarization, language diarization, code-mixing, conversational speech, DISPLACE challenge.

\section{Introduction}
\label{sec:DISPLACE23_intro}
In multilingual communities, social conversations frequently involve code-mixed and code-switched speech with multiple speakers~\cite{2009cambridgeBookBullock, 2017IEEEASRUYilmaz}.  \emph{Code-mixing} is the scenario where words or morphemes from one language (secondary) are used within a sentence context of another language, and \emph{code-switching} is the scenario where a language switch happens at the sentence or phrase level. Both code-mixing and code-switching are prevalent in multilingual communities like Asia, USA, and Europe~\cite{1998RoutledgeBookAuer, 2018SLTUSpoorthy,2013IEEECSICSIPLyu}.  As most of the current speech technologies are developed for single-speaker monolingual settings, the key first step in dealing with code-switched multi-speaker recordings would be the segregation of the input stream based on speaker and language.

The  term \emph{Diarization} was initially associated with the task of 
detecting and segmenting homogeneous audio regions based on speaker identity. This task,  widely known as speaker diarization (SD), generates the answer for ``who spoke when". In the past few years, the term diarization has also been  used in linguistic context. 
Language diarization (LD) is the task of identifying ``which language was spoken when", and aims to cluster segments of the same language~\cite{2013ICASSPLyu, 2021InterspeehLiu}. 

Most conventional speech processing systems, such as automatic speech recognition (ASR) and SD, are developed for monolingual scenarios.  At the same time, language recognition systems are not benchmarked  in multi-speaker settings. 
Further, most of the previous dataset building efforts focused on either of these challenges in isolation. 

In this paper, we describe the details of the Interspeech DIarization of SPeaker and LAnguage in Conversational Environments (DISPLACE) challenge. The key contributions of this challenge are:
\begin{itemize}
    \item Build a  dataset resource of multi-lingual multi-speaker conversational speech, with code switching, natural overlaps, reverberation and noise, along with speech, speaker and language annotations. 
    \item Benchmark speaker and language diarization on this dataset and announce an open call for system development.
    \item Provide a baseline system which is made available to the challenge participants, that performs speech activity detection, speaker diarization and language diarization.  
    \item A leader-board style evaluation platform for technology development. 
\end{itemize}

 

\begin{figure*}[t!]
\centerline{
\includegraphics[width=0.33\textwidth, height = 0.29\textwidth]{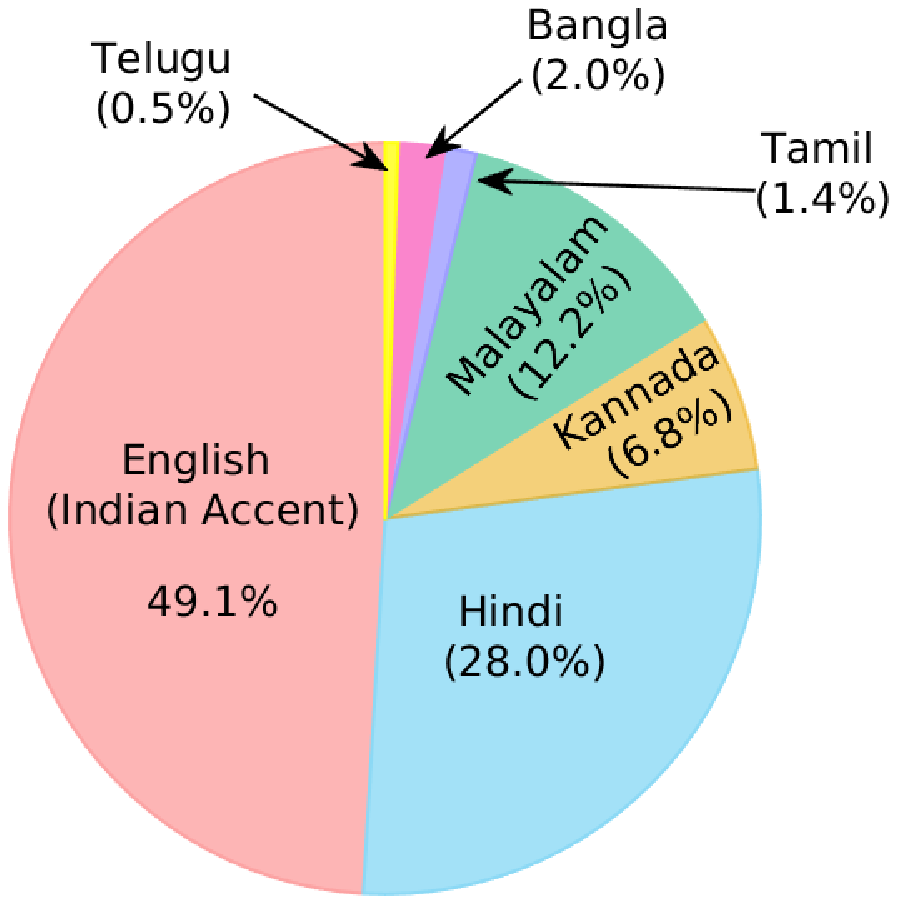}
\includegraphics[width=0.37\textwidth, height = 0.29\textwidth]{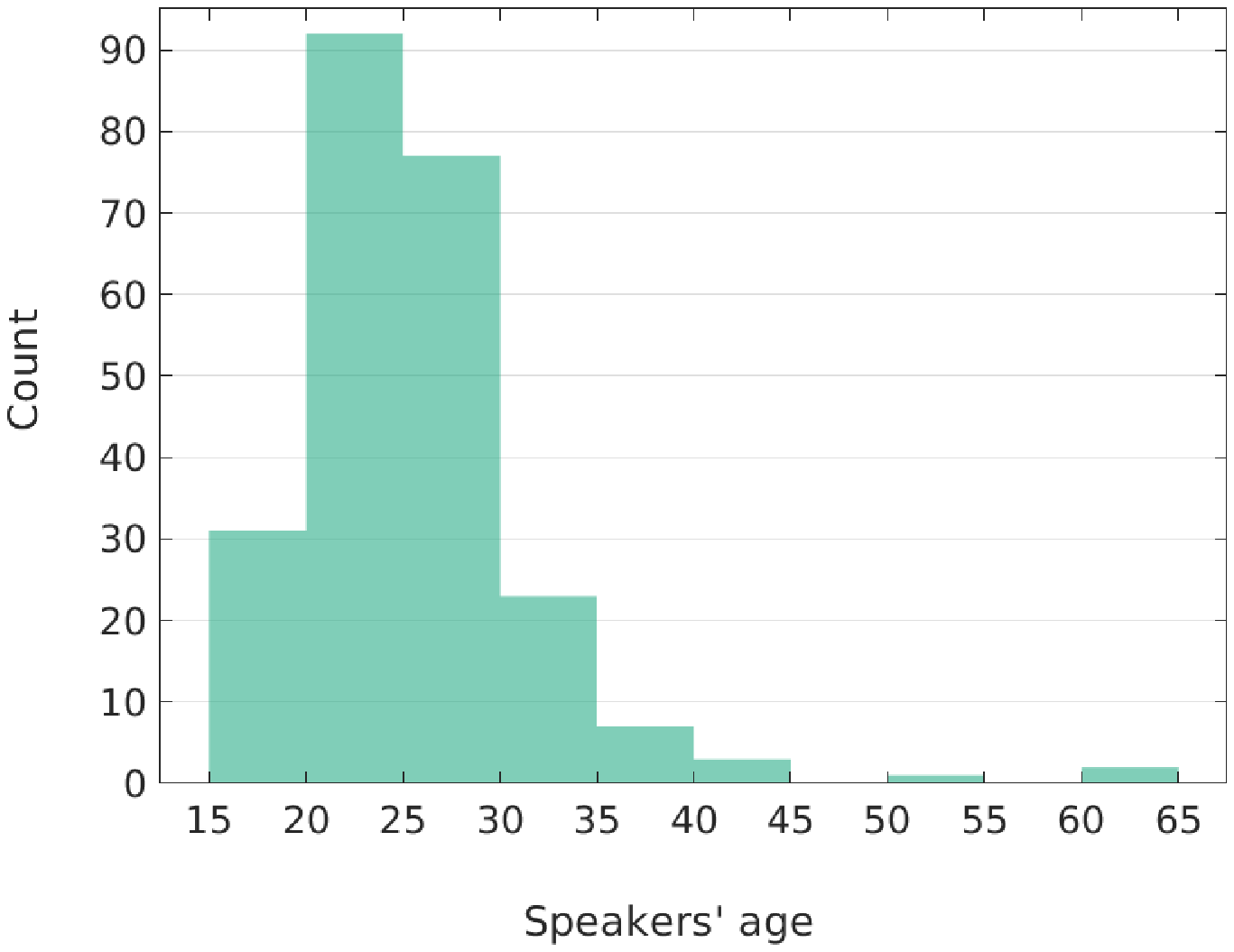}
\hspace{-0.2 cm}
\includegraphics[width=0.36\textwidth]{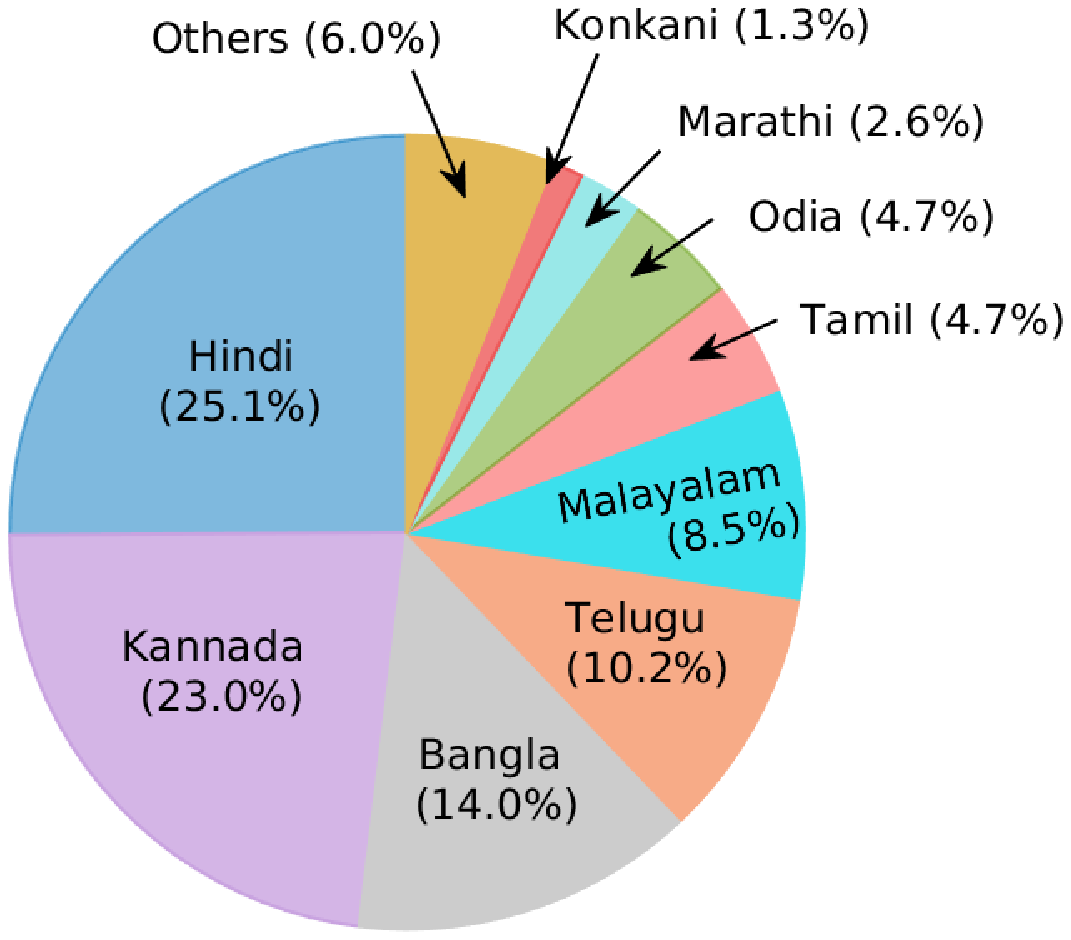}
}
\centerline{
(a)
\hspace{0.3\textwidth}(b) 	\hspace{0.35\textwidth}(c)
}
\caption{The DISPLACE corpus statistics of (a) Spoken language, (b) Speaker  age, and (c) Primary language (L1) of the speakers.}
\label{fig:DISPLACE_corpus_stats}

\end{figure*}
\section{Related Work}
\label{sec:DISPLACE23_relatedWorks}
\textbf{Speaker Diarization} - The majority of the early research in SD was driven by the National Institute of Standards and Technology-Rich Transcription (NIST-RT)~\cite{NISTRT,2004RT} evaluations on Broadcast News (BN) and conversational telephonic speech in English.  The  RT-05S~\cite{2005RT05SFiscus}, RT-06S, RT-07S~\cite{2007RT07Fiscus} and RT-09 evaluations considered conference and lecture room meeting domain (in English) for the SD task~\cite{NISTRT}. 
The NIST-RT evaluations also contributed the Diarization Error Rate (DER), which remains the primary evaluation metric for the SD systems.  Recently, there have been several evaluation challenges, namely, the DIHARD challenge \cite{2018DIHARDIRyant,2019InterspeechDIHARD2Ryant,2021InterspeechDIHARD3Ryant}, Fearless Steps Series~\cite{2019InterspeechFearlessHansen, 2020InterspeechFearless2Joglekar}, the Iberspeech-RTVE~\cite{2019IberSpeechChallengeLleida}, CHiME-6~\cite{2020CHiMEWatanabe} and VoxSRC-20~\cite{2020VoxsrcNagrani}. 

\noindent \textbf{Language Diarization} - In recent years, LD has become one of the contemporary research issues in multilingual speech processing~\cite{2021InterspeehLiu, 2022ArXivLiu}. In~\cite{2013IEEECSICSIPLyu}, Lyu et al. attempted the LD task on South-East-Asia Mandarin/English (SEAME) dataset~\cite{2010InterspeehLyu}. 
Yilmaz et al.~\cite{2017IEEEASRUYilmaz} used a radio broadcast dataset containing Frisian-Dutch code-switching for LD task. The first attempt for LD in the Indian languages was reported by Spoorthy et al.~\cite{2018SLTUSpoorthy}. 
Recently, Shah et al.~\cite{2020WSTCSMCShah} organized a workshop that included the LID task in code-switched data. 
The data used in this workshop is referred to as WSTCSMC~\cite{2020WSTCSMCShah}. 



\section{DISPLACE Corpus}
\label{sec:DISPLACE23_Corpus}


\subsection{Recording Setup}
\label{subsec:DISPLACE23_recording_setup}
The data is recorded at two different academic institutes. 
The recording rooms deployed are different in shape, size, and acoustic properties. These recording rooms did not have any specific soundproof settings and hence, the recordings contain natural noise and background speech seen in outdoor data collection settings. 
The data collection paradigm contains a close-talking microphone worn by each speaker and a common far-ﬁeld desktop microphone. The close-talking audio was recorded using a lapel microphone connected to either an audio recorder  or an Android phone. 
The participants were seated in either a circular or semi-circular setting and the far-field omni-directional microphone was used to collect the conversational speech.  
Speakers were approximately equi-distant from each other and the far-field microphone. All the recordings contain single-channel data.
The worn microphone speech was only recorded for annotation purposes (for ease of human listening to determine the speech, speaker and language boundaries). All the system development and evaluation is carried out using the far-field microphone recordings.

\subsection{Data Collection}
\label{subsec:DISPLACE23_data_collection}

Each conversation of duration $30$-$60$  minutes comprises $3$-$5$ participants. The participants had self-reported proficiency in at least one Indian language (L1) along with Indian accented English.  The participants that came together for a conversation were selected based on the L1 languages. The topic of conversation was chosen among a diverse set of choices like  climate change, culture, politics, sports, and entertainment. Before the recording, the participants also provided written consent and were  informed of the  guidelines. The participants also received monetary compensation for the recording. Finally, all the worn-mic recordings are time-aligned  with the far-field audio, resampled at 16 kHz, and normalized to $[-1,1]$ range.



\begin{figure*}[t]
\centerline{
\includegraphics[width=0.325\textwidth]{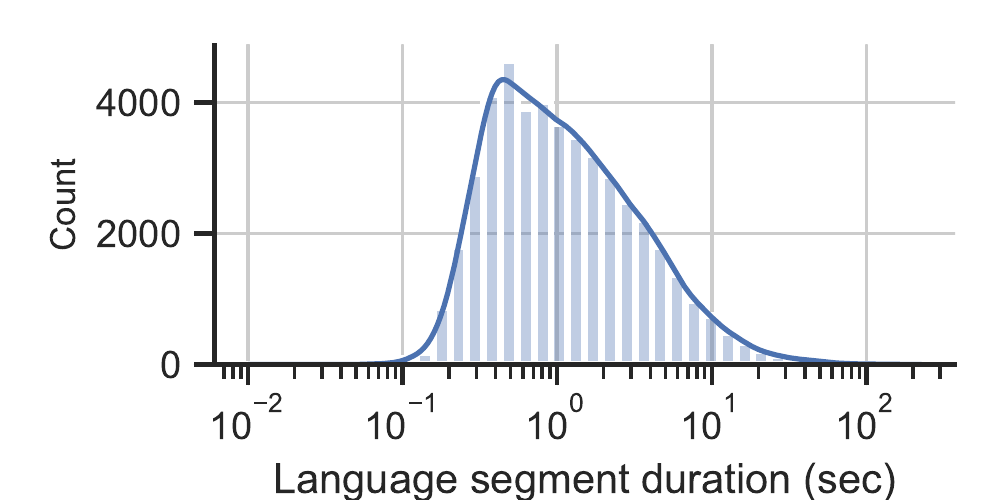}
\includegraphics[width=0.325\textwidth]{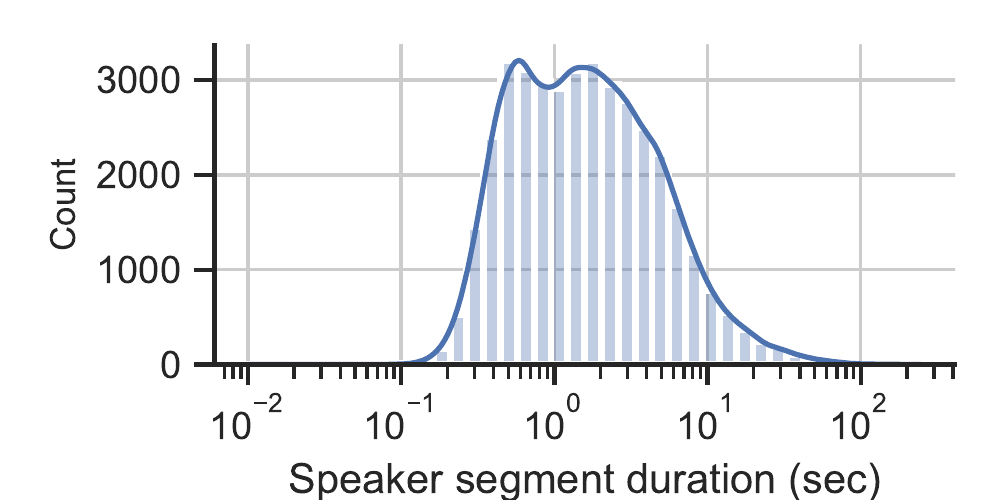}
\includegraphics[width=0.325\textwidth]{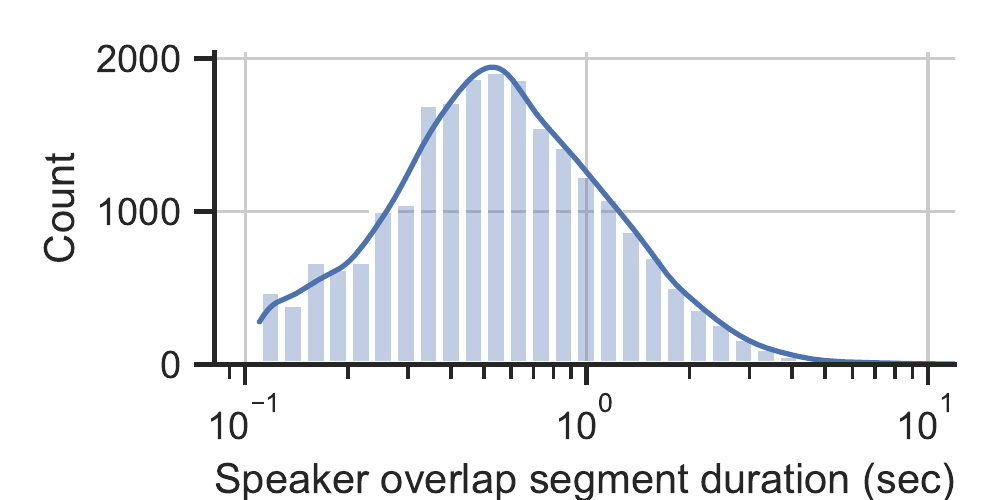}
}
\centerline{
(a)
\hspace{0.32\textwidth}(b) \hspace{0.32\textwidth} (c)
}
\caption{Histograms of (a) language, (b) speaker, and (c) overlap segment durations obtained from the DISPLACE corpus.}

\label{fig:DISPLACE_lan_spkr_seg_distri}
\end{figure*}

\subsection{Annotation}
\label{subsec:DISPLACE23_annotation}

The data annotations were generated with professional annotators, who listened to the   lapel microphone recordings. Each close-talking microphone is labeled  based on the target speaker (participant wearing the microphone). 

The first annotation task was to mark the speech activity of the target speaker (speaker activity). The rest of the regions  (including silence, long-pause, and  speech regions of non-target speakers) are marked as   \emph{non-speech}. 
 The non-speech audio of the target speaker, such as laughing, coughing, tongue clicks, etc., are annotated. The prominent background sounds, such as telephone rings, car honk, etc are also marked separately. 

The second annotation task was to mark language labels corresponding to the speaker activity regions. The language label was marked for every word to  incorporate code-mixing of short linguistic content. 

A third annotation task was  undertaken to generate the multilingual transcripts of the spoken content. These transcripts are not released as part of the DISPLACE challenge and will be used for future evaluations. 

The  participants use a lot of fillers and back-channel words, such as \emph{ahh}, \emph{umm}, and \emph{ohh}. For such cases, past and future contexts across the non-lexical words or sounds were considered for assigning a language label. The annotation process was a tedious one  involving speech, speaker and language annotations. We also employed multiple levels of quality checks before arriving at the final annotations used in the challenge. 
The annotations obtained for all the participants' lapel microphones are combined to generate a single Rich Transcription Time Marked (RTTM) annotation file for each conversation.





\subsection{Development and Evaluation set}
\label{subsec:dev_eva_ldata}
For the challenge, systems are evaluated on single channel far-ﬁeld audio. Only the development and evaluation sets were released. There was no training data provided, and the participants were free to use any proprietary and/or public resource for training the diarization systems. The Development (Dev) and Evaluation (Eval) set consists of $\approx 15.5$ hours ($27$ recordings) and $16$ hours ($29$ recordings) of multilingual conversations, respectively. The evaluation was done in two phases, namely, Phase-1 and Phase-2. The Phase-1 evaluation set consists of a subset of the full evaluation set with $20$ recordings spanning $11.5$ hours, and the Phase-2 evaluation set consisted of the full Eval set.  The Phase-1 evaluation ran for $7$ weeks (from Jan-15-2023 till March-4-2023), while Phase-2 was open till May-22-2023.  

The speakers in the evaluation and development sets are mutually exclusive. Also, the Eval recordings   contain  new languages not seen in the Dev set. The DISPLACE corpus {\color{black}(DEV and Eval Phase-1)} contains conversations in Hindi, Kannada, Bengali, Malayalam, Telugu, Tamil, and Indian English. The distribution of these languages is illustrated in Figure~\ref{fig:DISPLACE_corpus_stats} (a). The corpus also contains a significant amount (16.56\%) of overlapped speech.  Majority of the speakers' age ranges from $17$ to $40$ years (Figure~\ref{fig:DISPLACE_corpus_stats} (b)). The dataset contains $33$\% female and $67$\% male participants. The speakers in the DISPLACE corpus had different L1 languages, such as Bengali, Hindi, Kannada, Konkani, Malayalam, Marathi, etc. Figure~\ref{fig:DISPLACE_corpus_stats} (c) displays the distribution  of L1 languages spoken by speakers in the DISPLACE dataset (Dev and Eval Phase-1). The \emph{Others} category includes L1 languages  like Assamese, Gujarati, Kashmiri, Maithili, Nepali, Punjabi and Tulu. The corpus has a variety of speech accents as well.   

Figure~\ref{fig:DISPLACE_lan_spkr_seg_distri} demonstrates the distribution of language, speaker, and overlap segment durations. A language/speaker segment represents a homogeneous speech region containing a single language/speaker. A segment containing simultaneous speech of multiple speakers is considered as a speaker overlap segment. In Figures~\ref{fig:DISPLACE_lan_spkr_seg_distri} (a), (b), and (c), x-axes represent the segment duration (s) and y-axes denote segment count. In  Figure~\ref{fig:DISPLACE_lan_spkr_seg_distri} (a), the majority ($99.87\%$) of the language turns have a duration in the range of $0.10$s  to $100$s. Similarly, $99.96\%$ of the speaker turn durations vary from $0.1$s to $100$s.  As shown in Figure~\ref{fig:DISPLACE_lan_spkr_seg_distri} (c), overlap segment durations vary from $0.10$s to $10$s. Approximately $90.60\%$ overlap segments lie between $0.2$s to $4$s duration. These language, speaker, and overlap turn statistics show that frequent speaker and language switching is common in informal social conversations.


\begin{figure*}[t!]
\centerline{
\includegraphics[width=0.45\textwidth]{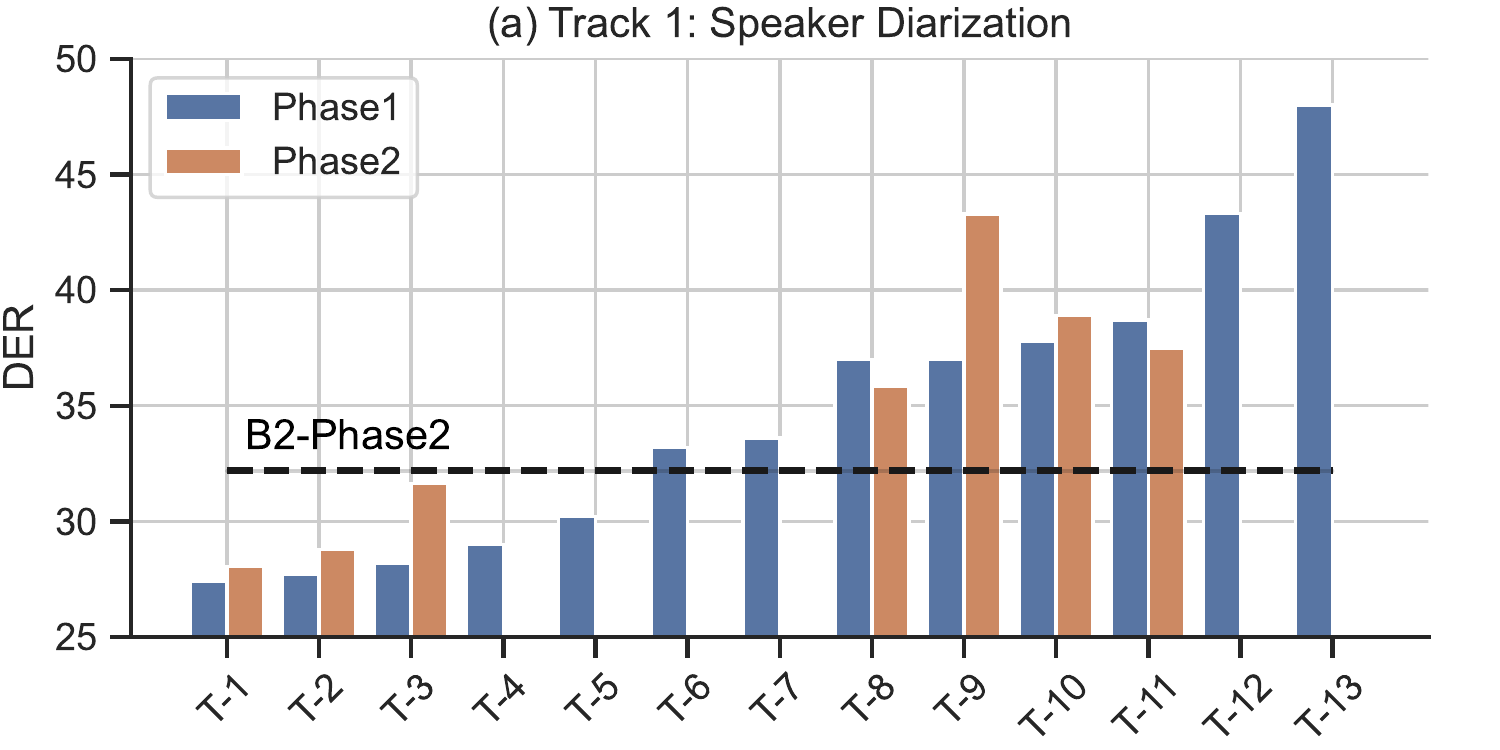}
\includegraphics[width=0.45\textwidth]{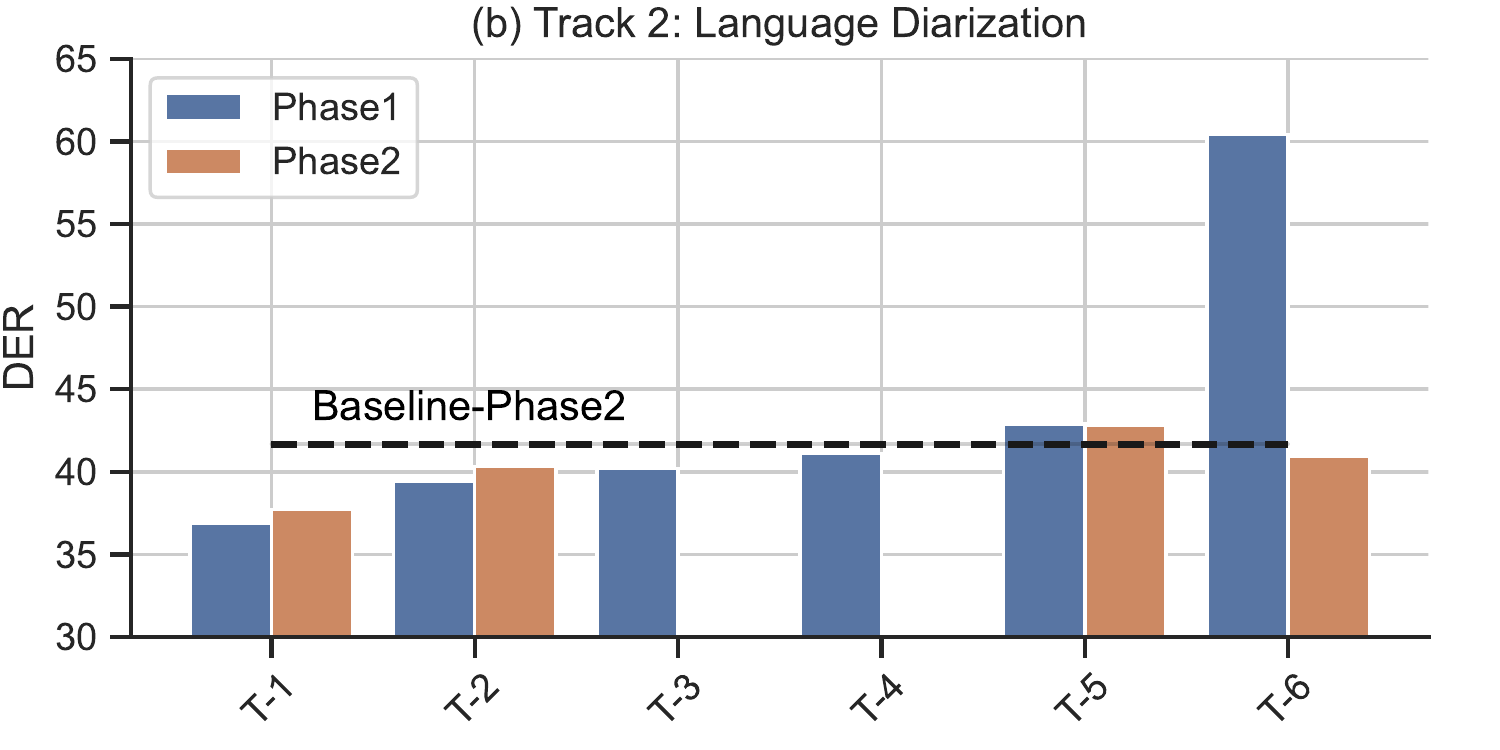}
}
\centerline{(a) \hspace{0.5\textwidth}(b)}

\caption{Eval Phase1 (blue bars) and Phase2 (orange bars) performance (DER $\%$) of all the participating teams for (a) speaker and (b) language diarization tracks.}
\label{fig:Eval_Phase1_DER}
\end{figure*}

\section{Challenge Tasks}
\label{sec:DISPLACE23_tasks}
The challenge tasks entail diarization of  each conversational audio based on speaker and language information. The speech activity detection is not provided and the participants are encouraged to develop their own system. 
Submissions to both tracks are evaluated only on speech-based speaker activity regions, including voiced back-channels and fillers. 
However, non-speech speaker activities, such as laughing, clapping, sneezing, etc., are excluded as non-speech. 
The DISPLACE challenge   contains two tracks, 
\begin{itemize}
\item 
 {{Track-1: Speaker Diarization in multilingual scenarios}}

\item 
 {{Track-2: Language Diarization in multi-speaker settings}}
\end{itemize} 

\noindent The  metric used for both tracks is the  diarization error rate (DER). 
\begin{equation}
    DER = \left(\frac{D_{FA} + D_{miss} + D_{error}}{D_{total}}\right) \times 100
\end{equation}
where, $D_{FA}$ represents the total speech  duration which is not attributed to a reference speaker/language, $D_{miss}$ denotes the total reference speaker/language duration, which is not attributed to a system speaker/language, $D_{error}$ represents the total system speaker/language duration attributed to the wrong reference speaker/language, and $D_{total}$ denotes the total reference speaker/language duration.
For the DISPLACE challenge, DER is computed with overlap and without collar. 

\section{Baseline Systems}
\label{sec:DISPLACE23_BS}

\subsection{Speech Activity Detection (SAD)}
\label{subsec:DISPLACE2023_SAD}
The SAD system from the DIHARD-III baseline~\cite{2021InterspeechDIHARD3Ryant} is used. The   model  contains $5$ Time Delay Neural Network (TDNN) layers~\cite{peddinti2015time} followed by 2 statistics pooling layers~\cite{ghahremani2016acoustic}. The DNN was trained on DIHARD-III development set to perform speech and non-speech classification and this model is trained for $40$ epochs. 







\begin{table}[t!]
\centering
\caption{Baseline speaker diarization results (in terms of DER \%) for the DISPLACE development (Dev.) and evaluation (Eval) data (Phase-1 and Phase-2) using  AHC and spectral clustering, followed by VB-HMM re-segmentation.}\label{tab:spk_diar_baseline_results}
\begin{tabular}{clccccc}
\toprule
                    &                  & \multicolumn{3}{c}{\textbf{DER (\%)}}  \\
                    \cmidrule(l){3-5} 
                    & \textbf{Method}           & \textbf{Dev}    & \multicolumn{2}{c}{\textbf{Eval}}  \\
                    \cmidrule(l){4-5} 
                    &   &     & \textbf{Phase1} & \textbf{Phase2}  \\
                    \cmidrule(l){1-5}  
\multirow{2}{*}{B1} & AHC              & 33.97  & 40.98   &  38.86 \\
                    & AHC+VB-HMM          & 32.57  & 40.08 &  37.96    \\
                     \cmidrule(l){1-5}
\multirow{2}{*}{B2} & SC & 31.66  & 33.58   &  32.51  \\
                    & SC+VB-HMM          & 31.17  & 32.94  & 32.18     \\
\bottomrule
\end{tabular}
\end{table}

\subsection{Speaker Diarization (SD)}
\label{subsec:DISPLACE23_BS_SD}

Speaker diarization system,   based on DIHARD-III~\cite{2021InterspeechDIHARD3Ryant}, is also adapted for this purpose. This modeling framework involves segmenting the audio into short overlapping segments of $1.5$s with $0.25$s shift and extracting speaker embeddings (x-vectors). These are then used to perform probabilistic linear discriminant analysis (PLDA) scoring followed by agglomerative
hierarchical clustering (AHC). For refining speaker boundaries, re-segmentation is performed using VB-HMM with posterior scaling~\cite{diez2018speaker,singh2019leap}.

The x-vector extractor model is  a $13$-layer Extended-Time Delay Neural Network (ETDNN)~\cite{snyder2019speaker}, which takes the $40$D mel-spectrogram features as input to extract $512$D x-vectors. The ETDNN model is trained on the VoxCeleb1~\cite{nagrani2017voxceleb} and VoxCeleb2~\cite{Chung2018} datasets, for speaker identification task, to discriminate among the $7,146$ speakers. A PLDA model is trained using x-vectors extracted from a subset of VoxCeleb1 and VoxCeleb2. The x-vectors are centered and whitened using the statistics estimated from the DISPLACE Dev data. The x-vectors are also unit-length normalized for PLDA training. 

We have also experimented with spectral clustering (SC)~\cite{2019InterspeechSCLin} algorithm along with AHC. The input to the SC are processed PLDA scores, obtained by applying sigmoid with temperature scaling of $0.1$.  
The VB-HMM is initialized separately for each recording from the clustering output. The hyper-parameter values for clustering algorithms and VB-HMM are set based on the performance  on the DISPLACE Dev set. Thus, we consider two baseline systems - AHC+VB-HMM (B1) and SC+VB-HMM (B2). The DER results are reported in Table~\ref{tab:spk_diar_baseline_results}. It can be observed that baseline system B2 outperforms B1 on Eval Phase1 and Phase2 data relatively by $18$\% and $15\%$, respectively. 
 




\begin{table}[t!]
\centering
\caption{Baseline language diarization results for the DISPLACE development and evaluation sets (Phase-1 and Phase-2) using  AHC and spectral clustering.}\label{tab:lang_diar_baseline_results}
\begin{tabular}{@{}lccc}
\toprule
\multirow{2}{*}{\textbf{Method}} & \multicolumn{3}{c}{DER (\%)} \\ 
\cmidrule(l){2-4} 
 & \multicolumn{1}{c}{\textbf{Dev}} & \multicolumn{2}{c}{\textbf{Eval}} \\ 
 \cmidrule(l){3-4} 
  & &\textbf{Phase1} & \textbf{Phase2} \\ 
 \midrule
AHC & \multicolumn{1}{c}{48.61} & 41.76 & 41.92 \\
Spec. Clustering & \multicolumn{1}{c}{48.48} & 41.48 & 41.67\\ \bottomrule
\end{tabular}
\end{table}

\subsection{Language Diarization (LD)}\label{subsec:DISPLACE23_BS_LD}
Existing models for language diarization assume that the test audio recordings contain two known languages \cite{2017IEEEASRUYilmaz}, as opposed to the DISPLACE dataset, which deals with a more generic case involving unknown languages. Hence, our baseline systems were redesigned for this task and used language embedding extraction at the segment level, followed by clustering.

We use the SpeechBrain  \cite{speechbrain} language identification model\footnote{\url{https://huggingface.co/speechbrain/lang-id-voxlingua107-ecapa}} based on the ECAPA-TDNN architecture \cite{desplanques2020ecapa} for embedding extraction. The model is trained on the Voxlingua107 dataset \cite{valk2021voxlingua107}, which contains $6628$ hours of speech from $107$ languages,  extracted from YouTube videos. We begin by segmenting the audio, containing only speech regions, into short overlapping segments of $0.4$s with $0.2$s shift. We extract $256$ dimensional embeddings at the segment level,  compute the pairwise cosine similarity score matrix, and apply a clustering algorithm to obtain the segment-level language predictions. We present the results of two clustering algorithms: agglomerative hierarchical clustering (AHC) and spectral clustering. This is given in Table~\ref{tab:lang_diar_baseline_results}.



\section{Challenge Results}
\label{sec:result}

Figure~\ref{fig:Eval_Phase1_DER} shows the Eval Phase-1 and Phase-2 DER performance of all the  participating teams for speaker and language diarization tracks. For SD track, a total of $13$ and $7$ teams participated in the Phase-1 and Phase-2 evaluations, respectively. The top three teams for SD track have reported significant improvements over baseline for both evaluation phases (as shown in Figure~\ref{fig:Eval_Phase1_DER} (a)). For Phase-1, the lowest DER obtained for the SD task is $27.4$\% (T-1) followed by $27.7$\% (T-2) and $28.2$\% (T-3). 
For Phase-2, the bottom three DER (for SD track) is $28.04$\% (T-1) followed by $28.79$\% (T-2) and $31.64$\% (T-3). The LD track (Track-2) received  six submissions in Phase-1, out of which four teams outperformed baseline DER. In Phase-2, only four teams participated in this track. 
For the LD track, DER for the top three performing teams are $36.9$ (T-1), $39.4$ (T-2) and $40.2$ (T-3), respectively for Phase-1. For Phase-2, the lowest DER obtained for the LD track is $37.72$ (T-1) followed by $40.32$ (T-2) and $41.67$ (Baseline).


\section{Summary}
In this paper, we have detailed the Interspeech DISPLACE challenge aimed at fostering research on processing multi-lingual multi-speaker conversational audio. The challenge entails two tracks, i) speaker diarization and ii) language diarization. The   dataset is common for the evaluation of both the tasks. The paper describes the data collection and annotation process.  

 Most of the previous SD challenges and existing works focus on monolingual scenarios, where a recording contains only one language. 
 Similarly. most of the previous works on LD used single speaker recordings. 
 To the best of our knowledge, no other publicly available dataset contains the plethora of diversity observed in the DISPLACE dataset in terms of multi-lingual, code-mixed, multi-speaker, natural conversational speech. 
 
In this paper, we also describe the baseline system and summarize the Phase-1 and Phase-2 submissions made by the challenge participants. Even with the  global scale of system development efforts, the dataset is challenging with the best results reporting a DER of $28.04$\% (Phase-2) for SD and $37.72$\% (Phase-2) for LD. These results   highlight the need for more future work  to enable  speech technology development in natural multi-lingual conversations.


\bibliographystyle{IEEEtran}
\bibliography{IS_23}

\end{document}